\documentclass[a4paper,twocolumn]{Gaia2004} % European paper size
\usepackage{times}      % for font
\usepackage{epsfig}     % for figure inclusion
\usepackage{natbib}     % for bibliography
\def\mn{MNRAS}
\def\apj{ApJ}
\def\apjs{ApJS}
\def\aap{A\&A}
\def\aj{AJ}
\def\b#1{{\bf#1}}\def\d{{\rm d}}\def\e{{\rm e}}\def\i{{\rm i}}
\renewcommand{\[}{\begin{equation}}
\renewcommand{\]}{\end{equation}}
\title{Modelling the Galaxy for GAIA}

\author{James Binney}
\affil{Oxford University, Rudolf Peierls Centre for Theoretical Physics, 1 Keble Road, Oxford OX1 3NP,
England}

\bibpunct{(}{)}{;}{a}{}{,}  % to set bibliography punctuation to A&A style

\begin{document}

\keywords{Galaxy models; stellar dynamics}

\maketitle

\begin{abstract}
Techniques for the construction of dynamical Galaxy models
should be considered essential infrastructure that should be put in place
before GAIA flies. Three possible modelling techniques are discussed.
Although one of these seems to have significantly more potential than the
other two, at this stage work should be done on all three.

A major effort is needed to decide how to make a  model consistent with a
catalogue such as that which GAIA will produce. Given the complexity of the
problem, it is argued that a hierarchy of models should be constructed, of
ever increasing complexity and quality of fit to the data. The
potential that resonances and tidal streams have to indicate how a model
should be refined is briefly discussed.

\end{abstract}

\section{Introduction}

A central goal of the GAIA mission is to teach us how the Galaxy functions
and how it was assembled. We can only claim to understand the structure of
the Galaxy when  we have a dynamical model galaxy that reproduces the
data. Therefore the construction of a satisfactory dynamical model is in a
sense a primary goal of the GAIA mission, for this model will encapsulate
the understanding of galactic structure that we have gleaned from GAIA.

Preliminary working models that are precursors of the final model will also
be essential tools as we endeavour to make astrophysical sense of the GAIA
catalogue.  Consequently, before launch we need to develop a model-building
capability, and with it produce dynamical models that reflect fairly fully
our current state of knowledge.

\section{Current status of Galaxy modelling}

The modern era of Galaxy models started in 1980, when the first version of
the Bahcall-Soneira model appeared \citep{BahcallS1}. This model broke new
ground by assuming that the Galaxy is built up of components like those seen
in external galaxies. Earlier work had centred on attempts to infer
three-dimensional stellar densities by directly inverting the observed star
counts. However, the solutions to the star-count equations are excessively
sensitive to errors in the assumed obscuration and the measured magnitudes,
so in practice it is essential to use the assumption that our Galaxy is
similar to external galaxies to choose between the infinity of statistically
equivalent solutions to the star-count equations. Bahcall \& Soneira showed
that a model inspired by data for external galaxies that had only a dozen or
so free parameters could reproduce the available star counts.

\cite{BahcallS1} did not consider kinematic data, but \cite{CaldwellO}
updated the classical work on mass models by fitting largely kinematic data
to a mass model that comprised a series of components like those seen in
external galaxies. These data included the Oort constants, the tangent-velocity
curve, the escape velocity at the Sun and the surface density of the disk
near the Sun.

\cite{BienaymeRC} were the first to fit both kinematic and star-count data
to a model of the Galaxy that was inspired by observations of external
galaxies. They broke the disk down into seven sub-populations by age. Then
they assumed that motion perpendicular to the plane is perfectly decoupled
from motion within the plane, and further assumed that as regards vertical
motion, each subpopulation is an isothermal component, with the velocity
dispersion determined by the observationally determined age-velocity
dispersion relation of disk stars.  Each sub-population was assumed to form
a disk of given functional form, and the thickness of the disk was
determined from the approximate formula
$\rho(R,z)/\rho(R,0)=\exp\{[\Phi(R,0)-\Phi(R,z)]/\sigma^2\}$, where $\Phi$
is an estimate of the overall Galactic potential. Once the thicknesses of
the sub-disks have been determined, the mass of the bulge and the parameters
of the dark halo were adjusted to ensure continued satisfaction of the
constraints on the rotation curve $v_c(R)$. Then the overall potential is
recalculated, and the disk thicknesses were redetermined in the new
potential. This cycle was continued until changes between iterations were
small.  The procedure was repeated several times, each time with a different
dark-matter disk arbitrarily superposed on the observed stellar disks. The
geometry and mass of this disk were fixed during the interations of the
potential. Star counts were used to discriminate between these dark-matter
disks; it turned out that the best fit to the star counts was obtained with
negligible mass in the dark-matter disk. Although in its essentials the
current `Besan\c on model' \citep{Robin03} is unchanged from the original
one, many refinements and extensions to have been made. In particular, the
current model fits near IR star counts and predicts proper motions and
radial velocities. It has a triaxial bulge and a warped, flaring disk. Its
big weakness is the assumption of constant velocity dispersions and
streaming velocities in the bulge and the stellar halo, and the neglect of
the non-axisymmetric component of the Galaxy's gravitational field.

A consensus that ours is a barred galaxy formed in the early 1990s
\citep{BlitzS1,Binneyetal} and models of the bulge/bar started to appear soon
after. \cite{BinneyGS} and \cite{Freudenreich} modelled the luminosity
density that is implied by the IR data from the COBE mission, while
\cite{Zhao1} and \cite{Hafneretal} used extensions of Schwarzschild's (1979)
modelling technique to produce dynamical models of the bar that predicted
proper motions in addition to being compatible with the COBE data. There was
an urgent need for such models to understand the data produced by searches
for microlensing events in fields near the Galactic centre. The interplay
between these data and Galaxy models makes rather a confusing story because
it has proved hard to estimate the errors on the optical depth to
microlensing in a given field.

The recent work of the Basel group \citep{Bissantz0,Bissantz1,Bissantz2} and the
microlensing collaborations \citep{Eros,popowski} seems at last to have
produced a reasonably coherent picture. \cite{Bissantz1} fit a model to
structures that are seen in the $(l,v)$ diagrams that one constructs from
spectral-line observations of HI and CO. The model is based on
hydrodynamical simulations of the flow of gas in the gravitational potential
of a density model that was fitted to the COBE data
\citep{Bissantz0}.  They show that structures observed in the $(l,v)$ plane
can be reproduced if three conditions are fulfilled: (a) the pattern speed
of the bar is assigned a value that is consistent with the one obtained by
\cite{DehnenPattern} from local stellar kinematics; (b) there are four spiral arms
(two weak, two strong) and they rotate at a much lower pattern speed; (c)
virtually all the mass inside the Sun is assigned to the stars rather than a
dark halo.  

\cite{Bissantz2} go on to construct a stellar-dynamical model that
reproduces the luminosity density inferred by \citep{Bissantz0}.  The model,
which has no free parameters, reproduces both (a) the stellar kinematics in
windows on the bulge, and (b) the microlensing event timescale distribution
determined by the MACHO collaboration \citep{Alcocketal}. The magnitude of
the microlensing optical depth towards bulge fields is still controversial,
but the latest results agree extremely well with the values predicted by
Bissantz \& Gerhard: in units of $10^{-6}$, the EROS collaboration report
optical depth $\tau_6=0.94\pm0.3$ at $(l,b)=(2.5^\circ,-4^\circ)$
\citep{Eros} while Bissantz \& Gerhard predicted $\tau_6=1.2$ at this
location; the MACHO collaboration report $\tau_6=2.17^{+0.47}_{-0.38}$ at
$(l,b)=(1.5^\circ,-2.68^\circ)$ \citep{popowski}, while Bissantz \& Gerhard
predicted $\tau_6=2.4$ at this location.

Thus there is now a body of evidence to suggest that the Galaxy's mass is
dominated by stars that can be traced by IR light rather than by invisible
objects such as WIMPS, and that dynamical galaxy models can successfully
integrate data from the entire spectrum of observational probes of the Milky
Way.

\section{Where do we go from here?}

Since 1980 there has been a steady increase in the extent to which Galaxy
models are dynamical. A model must predict stellar velocities if it is to
confront proper-motion and radial velocity data, or predict microlensing
timescale distributions, and it needs to predict the time-dependent,
non-axisymmetric gravitatinal potential in order to confront spectra-line
data for HI and CO. Some progress can be made by adopting characteristic
velocity dispersions for different stellar populations, but this is a very
poor expedient for several reasons. (a) Without a dynamical model, we do not
know how  the orientation of the velocity ellipsoid changes from place to
place. (b) It is not expected that any population will have Gaussian velocity
distributions, and a dynamical model is needed to predict how the
distributions depart from Gaussianity. (c) An arbitrarily chosen set of
velocity distributions at different locations for a given component are
guaranteed to be dynamically inconsistent. Therefore it is imperative that
we move to fully dynamical galaxy models. The question is simply, what
technology is most promising in this connection?

\subsection{Schwarzschild modelling}

The market for models of external galaxies is currently dominated by models
of the type pioneered by \cite{Schwarzschild}. One guesses the galactic
potential and calculates a few thousand judiciously chosen orbits in it,
keeping a record of how each orbit contributes to the observables, such as
the space density, surface brightness, mean-streaming velocity, or velocity
dispersion at a grid of points that covers the galaxy. Then one uses linear
or quadratic programming to find non-negative weights $w_i$ for each orbit
in the library such that the observations are well fitted by a model in
which a fraction  $w_i$ of the total mass is on the $i$th orbit.

Schwarzschild's technique has been used to construct spherical, axisymmetric
and triaxial galaxy models that fit a variety of observational constraints.
Thus it is a tried-and-tested technology of great flexibility.

It does have significant drawbacks, however. First the choice of initial
conditions from which to calculate orbits is at once important and obscure,
especially when the potential has a complex geometry, as the Galactic
potential has.  Second, different investigators will choose different
initial conditions and therefore obtain different orbits even when using the
same potential. So there is no straightforward way of comparing the
distributon functions of their models. Third, the method is computationally
very intensive because large numbers of phase-space locations have to be
stored for each of orbit.  Finally, predictions of the model are subject to
discreteness noise that is larger than one might naively suppose because
orbital densities tend to be cusped (and formally singular) at their edges
and there is no natural procedure for smoothing out these singularities.

\subsection{Torus modelling}

In Oxford over a number of years we developed a technique in which orbits
are not obtained as the time sequence that results from integration of the
equations of motion, but as images under a canonical map of an orbital torus
of the isochrone potential. Each orbit is specified by its actions $\b J$
and is represented by the coefficients $S_{\b n}(\b J')$ that define the
function $S(\b J',\theta)=\b J'\cdot\theta+\sum_{\b n}S_\b n\e^{\i\b
n\cdot\theta}$ that generates the map. Once the $S_\b n$ have been
determined, analytic expressions are available for $\b x(\theta)$ and $\b
v(\theta)$, so one can readily determine the velocity at which the orbit
would pass through any given location.  Since orbits are labelled by actions,
which define a true mapping of phase space, it is straightforward to
construct an orbit library by systematically sampling phase space at the
nodes of a regular grid of actions $\b J'$. Moreover, a good approximation
to an arbitrary orbit can be obtained by interpolating the $S_\b n(\b J')$.

If the orbit library is generated by torus mapping, it is easy to determine
the distributon function from the weights. When the orbit weigts are
normalized such that $\sum_iw_i=1$, and the distribution function is
normalized such that $\int\d^3\b x\d^3\b v\,f=1$, then
 \[
f(\b J)={1\over(2\pi)^3}\sum_iw_i\delta^{(3)}(\b J-\b J_i).
\]
 If the action-space gid is regular with spacing $\Delta$, we can obtain an
equivalent smoothed distribution function by replacing $\delta^{(3)}(\b J-\b J_i)$
by $\Delta^{-3}$ if $\b J$ lies within a cube of side $\Delta$ centred on
$\b J_i$, and zero otherwise.  Different modellers can easily compare their
smoothed distribution functions.  Finally, with torus mapping many fewer
numbers need to be stored for each orbit -- just the $S_\b n$ rather than
thousands of phase-space locations $(\b x,\b v)$. 

The drawbacks of torus mapping are these. First, it requires complex
special-purpose software, whereas orbit integration is trivial. Second, it
has to date only been demonstrated for systems that have two degrees of
freedom, such as an axisymmetric potential \citep{McGillB}, or a planar bar
\citep{kaasalainen1}. Finally, orbits are in a fictitious integrable
Hamiltonian \citep{kaasalainenB} rather than in the, probably non-integrable,
potential of interest. I return to this point below.

\subsection{Syer--Tremaine modelling}

In both the Schwarzschild and torus modelling strategies one starts by
calculating an orbit library, and the weights of orbits are determined only
after this step is complete.  \cite{SyerT} suggested an alternative stratey,
in which the weights are determined simultaneously with the integration of
the orbits. Combining these two steps reduces the large overhead involved in
storing large numbers of phase-space coordinates for individual orbits.
Moreover, with the Syer--Tremaine technique the potential does not have to
be fixed, but can be allowed to evolve in time, for example through the usual
self-consistency condition  of an N-body simulation.

To describe the Syer--Tremaine algorithm we need to define some notation.
Let $\b z\equiv(\b x,\b v)$ denote an arbitrary point in phase space. Then
each observable $y_\alpha$ is defined by a kernel $K_\alpha(\b z)$ through
 \[
y_\alpha=\int\d^6\b z\,f(\b z)K_\alpha(\b z).
\]
 For example, if $Y_\alpha$ is the density at some point $\b x_\alpha$, then
$K_\alpha(\b x,\b v)$ would be $\delta^{(3)}(\b x-\b x_\alpha)$. In an orbit model
we take $f$ to be of the form $f(\b z)=\sum_iw_i\delta^{(6)}(\b z-\b z_i)$ and the
integral in the last equation becomes a sum over orbits:
\[\label{discy}
y_\alpha=\sum_iw_iK_\alpha(\b z_i).
\]

If we simultaneously integrate a large number of orbits in a common
potential $\Phi$ (which might be the time-dependent potential that is
obtained by assigning each particle a mass $w_iM$), then through equation
(\ref{discy}) each observable becomes a function of time. Let $Y_\alpha$ be
the required value of this observable,  then Syer \& Tremaine adjust the
value of the weight of the $i$th orbit at a rate
 \[
{\d w_i\over\d t}=- w_i\sum_\alpha{K_\alpha[\b z_i(t)]\over Z_\alpha}
\left({y_\alpha\over Y_\alpha}-1\right).
\]
 Here the positive numbers $Z_\alpha$ are chosen judiciously to stress the
importance of satisfying particular constraints, and can be increased to
slow the rate at which the weights are adjusted. The numerator $K_\alpha[\b
z_i(t)]$ ensures that a discrepancy between $y_\alpha(t)$ and $Y_\alpha$
impacts $w_i$ only in so far as the orbit contributes to $y_\alpha$. The right
side starts with a minus sign to ensure that $w_i$ is decreased if
$y_\alpha>Y_\alpha$ and the orbit tends to increase $y_\alpha$.
\cite{Bissantz2} have recently demonstrated the value of the Syer \& Tremaine
algorithm by using it to construct a dynamical model of the inner Galaxy in
the pre-determined potential of \cite{Bissantz0}. 

N-body simulations have been enormously important for the development of our
understanding of galactic dynamics. To date they have been of rather limited
use in modelling specific galaxies, because the structure of an N-body model
has been determined in an obscure way by the initial conditions from which
it is started. In fact, a major motivation for developing other modelling
techniques has been the requirement for initial conditions that will lead to
N-body models that have a specified structure \citep[e.g][]{KuijkenD}.
Nothwithstanding this difficulty, \cite{Fux} was able to find an N-body
model that qualitatively fits observations of the inner Galaxy. It will be
interesting to see whether the Syer--Tremaine algorithm can be used to
refine a model like that of Fux until it matches all observational
constraints.

\section{Hierarchical modelling}

When trying to understand something that is complex, it is best to proceed
through a hierarchy of abstractions: first we paint a broad-bruish picture
that ignores many details. Then we look at areas in which our first picture
clearly conflicts with reality, and understand the reasons for this
conflict. Armed with this understanding, we refine our model to eliminate
these conflicts. Then we turn to the most important remaining areas of
disagreement between our model and reality, and so on. The process
terminates when we feel that we have nothing new or important to learn from
residual mismatches between theory and measurement.

This logic is nicely illustrated by the dynamics of the solar system. We
start from the model in which all planets move on Kepler ellipses around the
Sun. Then we consider the effect on planets such as the Earth of Jupiter's
gravitational field. To this point we have probably assumed that all bodies
lie in the ecliptic, and now we might consider the non-zero inclinations of
orbits. One by one we introduce disturbances caused by the masses of the
other planets. Then we might introduce corrections to the equations of motion from
general relativity, followed by consideration of effects that arise because
planets and moons are not point particles, but spinning non-spherical
bodies. As we proceed through this hierarchy of models, our orbits will
proceed from periodic, to quasi-periodic to chaotic. Models that we
ultimately reject as oversimplified will reveal structure that was
previously unsuspected, such as bands of unoccupied semi-major axes in the
asteroid belt. The chaos that we will ultimately have to confront  will be
understood in terms of resonances between the orbits we considered in the
previous level of abstraction. 

The impact of Hipparcos on our understanding of the dynamics of the solar
neighbourhood gives us a flavour of the complexity we will have to confront
in the GAIA catalogue. When the density of stars in the $(U,V)$ plane was
determined \citep{DehnenStruct,Fux}, it was found to be remarkably lumpy,
and the lumps contained old stars as well as young, so they could not be
just dissolving associations, as the classical interpretation of star
streams supposed.  Now that the radial velocities of the Hipparcos survey
stars are available, it has become clear that the Hyades-Pleiades and Sirius
moving groups are very heterogeous as regards age \citep{Famaey}. Evidently
these structures do not reflect the patchy nature of star formation, but
have a dynamical origin. They are probably generated by transient spiral
structure \citep{Desimone}, so they reflect departures of the Galaxy from
both axisymmetry and time-independence. Such structures will be most readily
understod by perturbing a steady-state, axisymmetric Galaxy model.

A model based on torus mapping is uniquely well suited to such a study because
its orbits are inherently quasi-periodic structures with known
angle-action coordinates. Consequently, we have everything we need to use
the  powerful techniques of canonical perturbation theory. 

 \begin{figure}[ht]
  \begin{center}
    \leavevmode
\centerline{\psfig{file=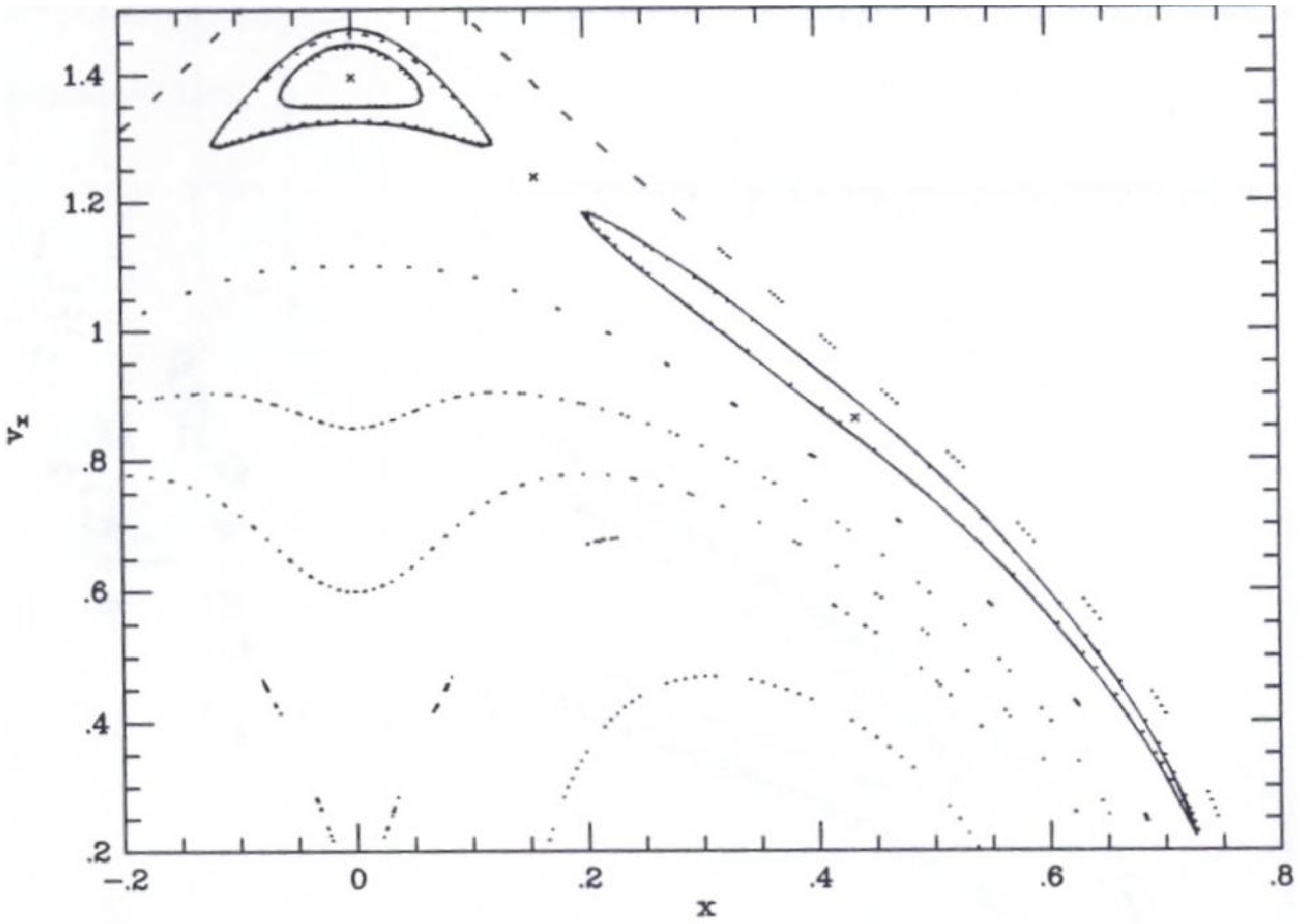,width=\hsize}}
\centerline{\psfig{file=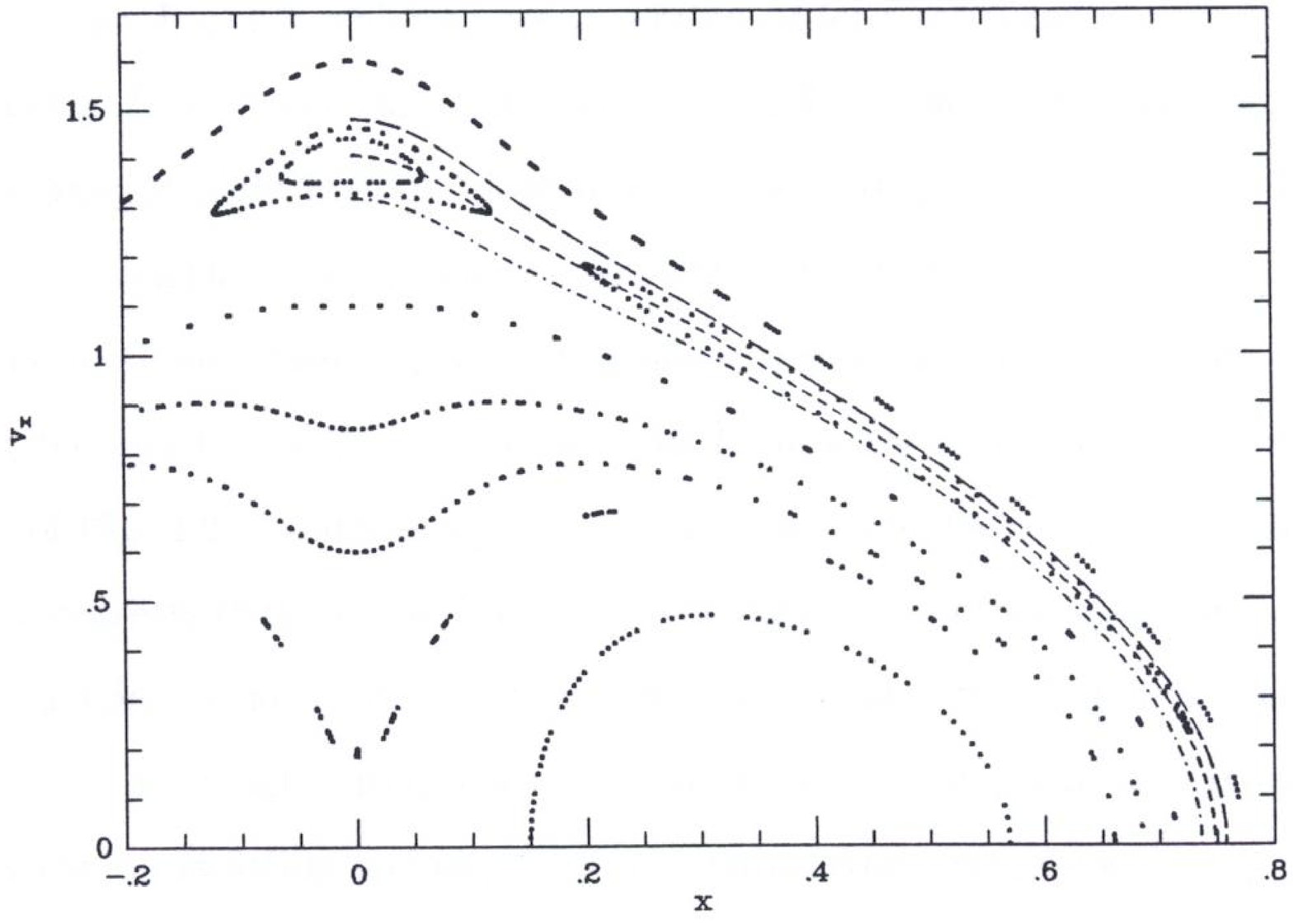,width=\hsize}}
   \end{center}
  \caption{Using perturbation theory to model a resonant family of orbits.
  Dots show consequents obtained by numerical integration. The curves in the
  top panel
  show resonant orbits obtained by applying perturbation theory to 
  orbits obtained by torus mapping. The curves in the lower panel show three
  of these orbits, one through the centre of the resonant region and one on
  each side. From \cite{KaasalainenT}.}
  \label{fig1}
\end{figure}

 \begin{figure}[ht]
  \begin{center}
    \leavevmode
 \centerline{\psfig{file=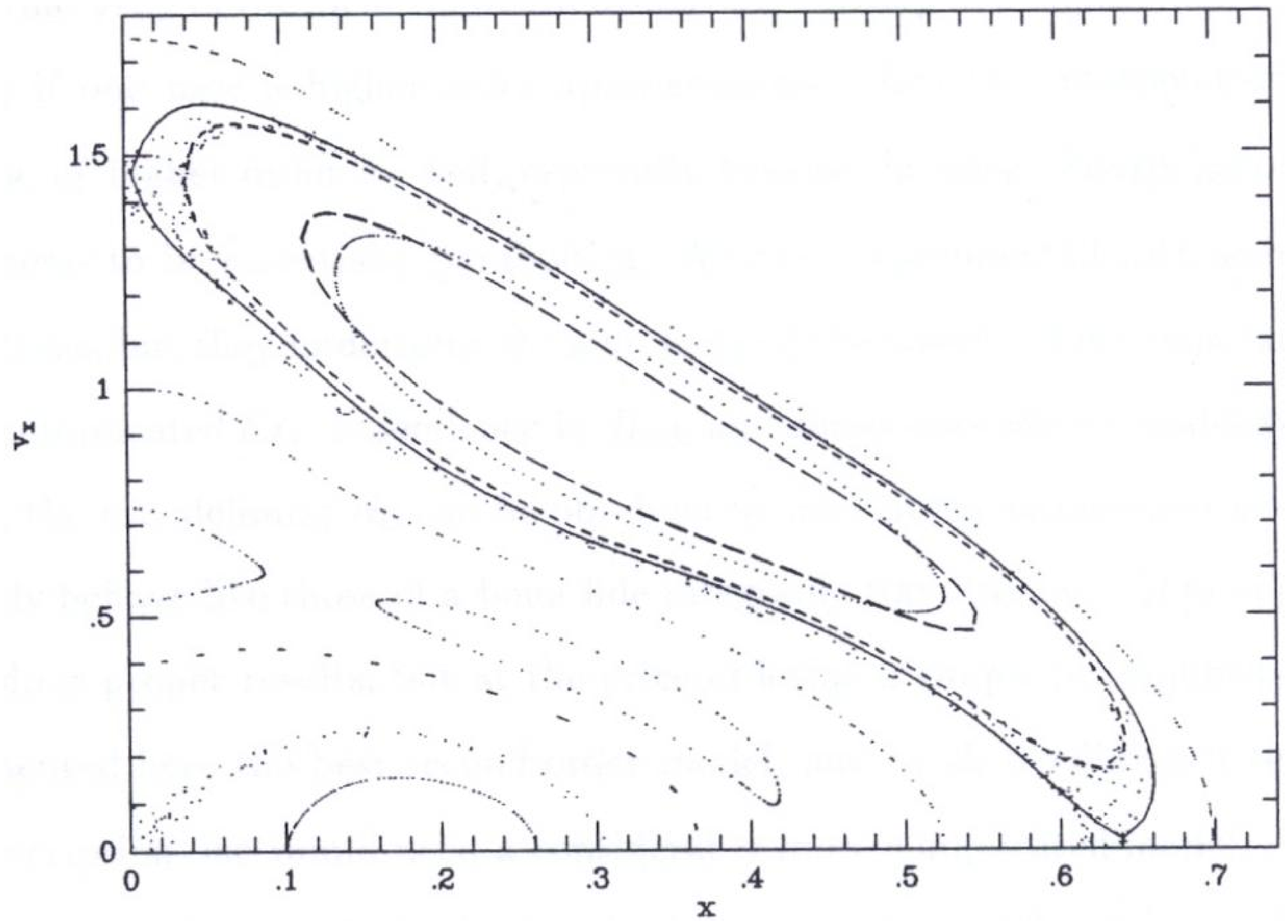,width=\hsize}}
 \centerline{\psfig{file=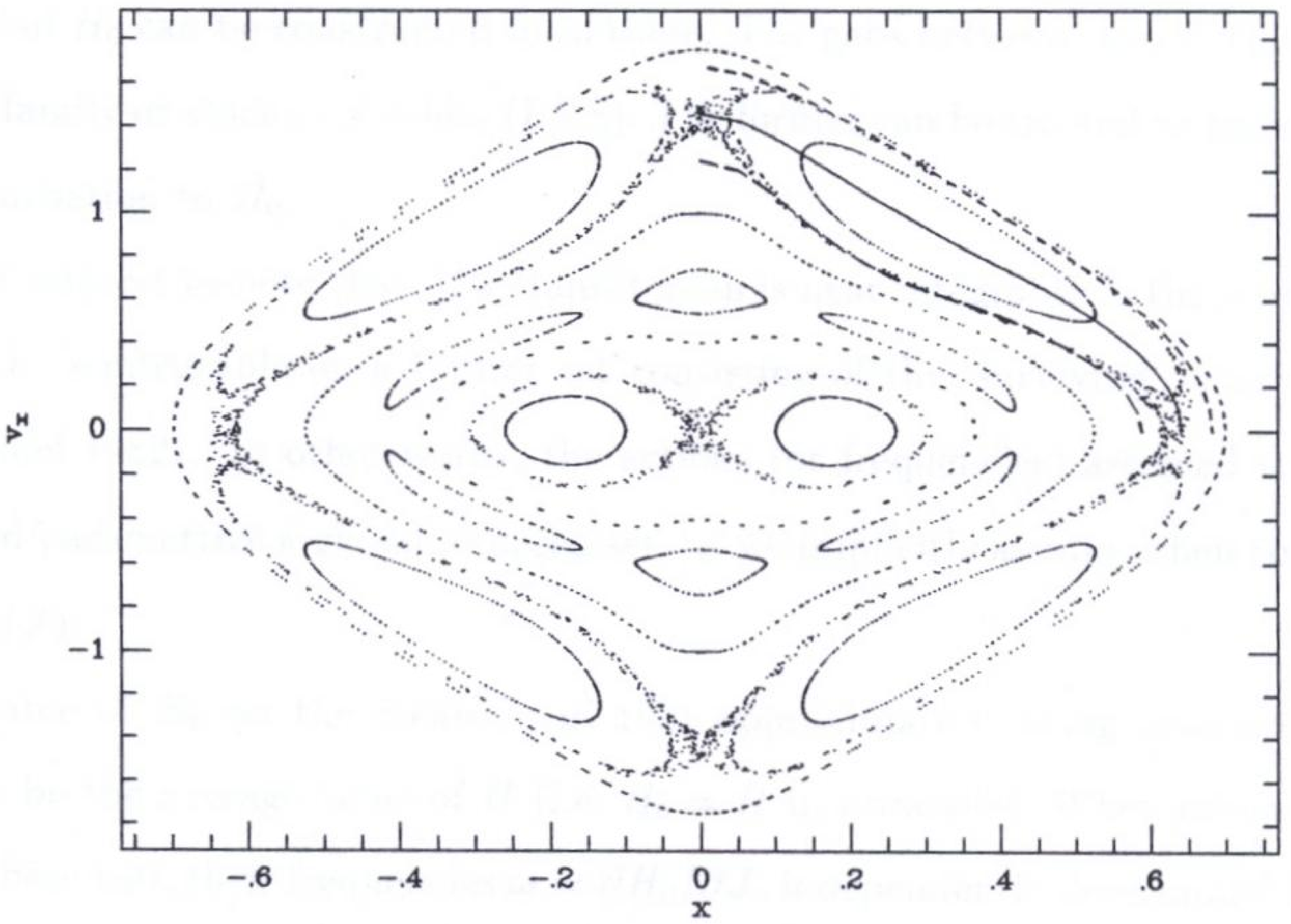,width=\hsize}}
   \end{center}
  \caption{Obtaining resonant orbits by direct torus mapping. The lower
  panel shows a surface of section that is largely taken up by several
  powerful  resonances. Three non-resonant orbits obtained by torus mapping
  are shown, one through the middle and one on each side of the largest
  resonant family. The uper panel shows several orbits of the resonant
  family that are obtained by directly mapping isochrone orbits. Notice that
  the chaotic region is nicely contained between two of these mapped orbits.
  From \cite{KaasalainenT}.}
  \label{fig2}
\end{figure}

Even in the absence of departures from axisymmetry or time-variation in the
potential, resonances between the three characteristic frequencies of a
quasi-periodic orbit can deform the orbital structure from that encountered
in analytically integrable potentials.  Important examples of this phenomenon
are encountered in the dynamics of triaxial elliptical galaxies, where
resonant `boxlets' almost entirely replace box orbits when the potential is
realistically cuspy \citep{MerrittF}, and in the dynamics of disk galaxies,
where the 1:1 resonance between radial and vertical oscillations probably
trapped significant numbers of thick-disk stars as the mass of the thin disk
built up \citep{SridharT}. \cite{kaasalainen3} has shown that such families
of resonant orbits may be very successfully modelled by applying
perturbation theory to orbits obtained by torus mapping.  If the resonant
family is exceptionally large, one may prefer to obtain its orbits directly
by torus mapping \citep{kaasalainen2} rather than through perturbation
theory. Figures \ref{fig1} and \ref{fig2} show examples of each approach to
a resonant family. Both figures show surfaces of section for motion in a
planar bar. In Figure \ref{fig1} a relatively weak resonance is successfuly
handled through perturbation theory, while in Figure \ref{fig2} a more
powerful resonance that induces significant chaos is handled by directly
mapping isochrone orbits into the resonant region. These examples
demonstrate that if we obtain orbits by torus mapping, we will be able to discover what
the Galaxy would look like in the absence of any particular resonant family
or chaotic region, so we will be able to ascribe particular features in the
data to particular resonances and chaotic zones. This facility will make the
modelling process more instructive than it would be if we adopted a simple
orbit-based technique.

\section{Confronting the data}

A dynamical model Galaxy  will consist of a gravitational potential $\Phi(\b
x)$ together with distribution functions $f_\alpha(\b J)$ for each of
several stellar populations. Each distribution function may be represented by
a set of orbital weights $w_i$, and the populations will consist of
probability distributions in mass $m$, metallicity $Z$ and age $\tau$ that a
star picked from the population has the specified characteristics. Thus a
Galaxy model will contain an extremely large number of parameters, and
fitting these to the data will be a formidable task.

Since so much of the Galaxy will be hidden from GAIA by dust, interpretation
of the GAIA catalogue will require a knowledge of the three-dimensional
distribution of dust. Such a model can be developed by the classical method
of comparing measured colours with the intrinsic colours of stars of known
spectral type and distance. At large distances from the Sun, even GAIA's
small parallax errors will give rise to significantly uncertain distances,
and these uncertainties will be an important limitation on the reliability
of any dust model that one builds in this way. 

Dynamical modelling offers the opportunity to refine our dust model because
Newton's laws of motion enable us to predict the luminosity density in
obscured regions from the densities and velocities that we see elsewhere,
and hence to detect obscuration without using colour data.
Moreover, they require that the luminosity distributions of hot components
are intrinsically smooth, so fluctuations in the star counts of these
populations at high spatial frequencies must arise from small scale
structure in the obscuring dust. Therefore, we should solve for the
distribution of dust at the same time as we are solving for the potential and
the orbital weights.

In principle one would like to fit a Galaxy model to the data by predicting
from the model the probability density $P(\alpha,\ldots,v)$ of detecting a
star at given values of the catalogue variables, such as celestial
coordinates $\alpha$, parallax $\varpi$, and proper motons $\mu$, and then
evaluating the likelihood $L=\prod_i P_i$, where the product runs over stars
in the catalogue and
 \begin{eqnarray}
P_i=\int\d^2\alpha\,\d\varpi\,\d^2\mu\,\d v\,
{\e^{-(\alpha-\alpha_i)^2/2\sigma^2_\alpha}\over2\pi\sigma^2_\alpha}
\nonumber\\
\times\cdots\times
{\e^{-(v-v_i)^2/2\sigma^2_v}\over(2\pi\sigma^2_v)^{1/2}}\,P(\alpha,\ldots,v)
\end{eqnarray}
 with $\alpha_i\ldots v_i$ the measured values and
$\sigma_\alpha\ldots\sigma_v$ the associated uncertainties.  Unfortunately,
it is likely to prove difficult to obtain the required probability density
$P$ from an orbit-based model, and we will be obliged to compare the real
catalogue to a pseudo-catalogue derived from the current model. Moreover,
standard optimization algorithms are unlikely to find the global maximum in
$L$ without significant astrophysical input from the modeller. In any event,
evaluating $P_i$ for each of $\sim10^9$ observed stars is a formidable
computational problem.  Devising efficient ways of fitting models to the
data clearly requires much more thought.

\subsection{Adaptive dynamics}

Fine structure in the Galaxy's phase space may provide crucial assistance in
fitting a model to the data. Two inevitable sources of fine structure are
(a) resonances, and (b) tidal streams. Resonances will sometimes be marked
by a sharp increase in the density of stars, as a consequence of resonant
trapping, while other resonances show a deficit of stars. Suppose the data
seem to require an enhanced density of stars at some point in action space
and you suspect that the enhancement is caused by a particular resonance. By
virtue of errors in the adopted potential $\Phi$, the frequencies will not
actually be in resonance at the centre of the enhancement. By appropriate
modification of $\Phi$ it will be straightforward to bring the frequencies
into resonance. By reducing the errors in the estimated actions of orbits, a
successful update of $\Phi$ will probably enhance the overdensity around the
resonance.  In fact, one might use the visibility of density enhancements to
adjust $\Phi$ very much as the visibility of stellar images is used with
adaptive optics to configure the telescope optics.

A tidal stream is a population of stars that are on very similar orbits --
the actions of the stars are narrowly distributed around the actions of the
orbit on which the dwarf galaxy or globular cluster was captured.
Consequently, in action space a tidal stream has higher contrast than it does
in real space, where the stars' diverging angle variables  gradually spread
the stars over the sky. Errors in $\Phi$ will tend to disperse a tidal
stream in action space, so again $\Phi$ can be tuned by making the tidal stream as
sharp a feature as possible.

\section{Conclusions}

Dynamical Galaxy models have a central role to play in attaining GAIA's core
goal of determining the structure and unravelling the history of the Milky
Way. 
Even though people have been building Galaxy models for over half a
century, we are still only beginning to construct fully dynamical models,
and we are very far from being able to build multi-component dynamical
models of the type that the GAIA will require.

At least three potentially viable Galaxy-modelling technologies can be
identified. One has been extensively used to model external galaxies, one
has the distinction of having been used to build the currently leading
Galaxy model, while the third technology is the least developed but
potentially the most powerful. At this point we would be wise to pursue all
three technologies.

Once constructed, a model needs to be confronted with the data. On account
of the important roles in this confrontation that will be played by
obscuration and parallax errors, there is no doubt in my mind that we need
to project the models into the space of GAIA's catalogue variables $(\alpha,
\varpi,\ldots)$. This projection is simple in principle, but will be
computationally intensive in practice.

The third and final task is to change the model to make it fit the data
better. This task is going to be extremely hard, and it is not clear at this
point what strategy we should adopt when addressing it. It seems possible
that features in the action-space density of stars associated with
resonances and tidal streams will help us to home in on the correct
potential. 

There is much to do and it is time we started doing it if we want to have a
reasonably complete box of tools in hand when the first data arrive in
2012--2013. The overall task is almost certainly too large for a single
institution to complete on its own, and the final galaxy-modelling machinery
ought to be at the disposal of the wider community than the dynamics
community since it will be required to evaluate the observable implications
of any change in the characteristics or kinematics of stars or interstellar
matter throughout the Galaxy.  Therefore, we should approach the problem
of building Galaxy models as an aspect of infrastructure work for GAIA,
rather than mere science exploitation.

I hope that in the course of the next year interested parties will enter
into discussions about how we might divide up the work, and define interface
standards that will enable the efforts of different groups to be combined in
different combinations. It is to be hoped that these discussions lead before
long to successful applications to funding bodies for the resources that
will be required to put the necessary infrastructure in place by 2012.

%\section*{Acknowledgments}


\begin{thebibliography}{}

\bibitem[Alcock et al.(2000)]{Alcocketal}
Alcock, C., et al., 2000, \apj, 541, 734

\bibitem[Afonso, et al.(2003)]{Eros}
Afonso, C., Albert J., Alard C., et al., 2003, \aap, 404, 145

\bibitem[Bahcall \& Soneira(1980)]{BahcallS1}
Bahcall, J.N. \& Soneira, R.M., 1980, \apjs, 44, 73

\bibitem[Bahcall \& Soneira(1984)]{BahcallS2}
Bahcall, J.N. \& Soneira, R.M., 1984, \apjs, 55, 67

\bibitem[Bienaym\'e, Robin \& Cr\'ez\'e(1987)]{BienaymeRC}
Bienaym\'e, O., Robin, A.C. \& Cr\'ez\'e, M., 1987, \aap, 180, 94

\bibitem[Binney et al.(1991)]{Binneyetal}
Binney, J.J., Gerhard, O.E., Stark, A.A., Bally, J. \& Uchida, K.I., 1991,
\mn, 252, 210

\bibitem[Binney, Gerhard \& Spegel(1997)]{BinneyGS}
Binney, J.J., Gerhard, O.E. \& Spegel, D.N., 1997, \mn, 288, 365

\bibitem[Bissantz, Debattista \& Gerhard(2004)]{Bissantz2}
Bissantz, N., Debattista, V.P. \& Gerhard, O.E., 2004, \apj, 601, L155

\bibitem[Bissantz, Englmaier \& Gerhard(2003)]{Bissantz1}
Bissantz, N., Englmaier, P. \& Gerhard, O.E., 2003, \mn, 340, 949

\bibitem[Bissantz \& Gerhard(2002)]{Bissantz0}
Bissantz, N. \& Gerhard, O.E., 2002, \mn, 330, 591

\bibitem[Blitz \& Spergel(1991)]{BlitzS1}
Blitz, L. \& Spergel, D.N., 1991, \apj, 379, 631

\bibitem[Caldwell \& Ostriker(1981)]{CaldwellO}
Caldwell, J.A.R. \& Ostriker, J.P., 1981, \apj, 251, 61

\bibitem[Dehnen(1998)]{DehnenStruct}
Dehnen, W., 1998, \aj, 115, 2384

\bibitem[Dehnen(2000)]{DehnenPattern}
Dehnen, W., 2000, \aj, 119, 800

\bibitem[De Simone, Wu \& Tremaine(2004)]{Desimone}
De Simone, R.S., Wu, X. \& Tremaine, S., 2004, \mn, 350, 627

\bibitem[Famaey et al.(2004)]{Famaey}
Famaey, B., Jorissen, A., Luri, X., Mayor, M., Udry, S., Dejonghe, H. \&
Turon, C., \aap, xx xx

\bibitem[Freudenreich(1998)]{Freudenreich}
Freudenreich, H.T., 1998, \apj, 492, 495

\bibitem[Fux(1997)]{Fux}
Fux, R., 1997, \aap, 327, 983

\bibitem[H\"afner et al.(2000)]{Hafneretal}
H\"afner, R., Evans, N.W., Dehnen, W. \& Binney, J., 2000, \mn, 314, 433

\bibitem[Kaasalainen(1994a)]{KaasalainenT}
Kaasalainen, M., 1994a, D.Phil thesis, Oxford University

\bibitem[Kaasalainen \& Binney(1994b)]{kaasalainen1}
Kaasalainen, M. \& Binney, J., 1994b, \mn, 268, 1041

\bibitem[Kaasalainen \& Binney(1994c)]{kaasalainenB}
Kaasalainen, M. \& Binney, J., 1994c, Phys.Rev.L., 73, 2377

\bibitem[Kaasalainen(1995a)]{kaasalainen2}
Kaasalainen, M., 1995a, \mn, 275, 162%closed

\bibitem[Kaasalainen(1995b)]{kaasalainen3}
Kaasalainen, M., 1995b, Phys.Rev.E., 52, 1193%chaos

\bibitem[Kuijken \& Dubinski(1995)]{KuijkenD}
Kuijken, K. \& Dubinski, J., 1995, \mn, 277, 1341

\bibitem[McGill \& Binney(1990)]{McGillB}
McGill, C. \& Binney, J., 1990, \mn, 244, 634

\bibitem[Merritt \& Fridman(1996)]{MerrittF}
Merritt, D. \& Fridman, T. 1996, \apj, 460, 136
 
\bibitem[Popowski et al.(2004)]{popowski}Popowski P., Griest K., Thomas
C., et al., 2004, ApJ, (astro-ph/0410319) 

\bibitem[Robin et al.(2003)]{Robin03}
Robin, A.C., Reyl\'e, C., Derri\'ere, S. \& Picaud, S., 2003, \aap, 409, 523

\bibitem[Schwarzschild(1979)]{Schwarzschild}
Schwarzschild, M., 1979, \apj, 232, 236

\bibitem[Sridhar \& Touma(1996)]{SridharT}
Sridhar, S. \& Touma, J. 1996, \mn, 279, 1263

\bibitem[Syer \& Tremaine(1996)]{SyerT}
Syer, D. \& Tremaine, S., 1996, \mn, 282, 223

\bibitem[Zhao(1996)]{Zhao1}
Zhao, H.-S., 1996, \mn, 283, 149

 \end{thebibliography}
\end{document}